\documentclass[aps,preprint,superscriptaddress]{revtex4-1}
\usepackage{graphicx}
\usepackage{amsmath}
\usepackage{makeidx}
\usepackage{amsfonts}
\usepackage{amssymb}

\begin{document}

\title{Bethe free-energy approximations for disordered quantum systems}

\author{I. Biazzo}
\email{indaco.biazzo@polito.it}
\affiliation{DISAT and Center for Computational Sciences, Politecnico di Torino, Corso Duca degli Abruzzi 24, 10129 Torino, Italy}

\author{A. Ramezanpour}
\email{aramezanpour@gmail.com}
\affiliation{DISAT and Center for Computational Sciences, Politecnico di Torino, Corso Duca degli Abruzzi 24, 10129 Torino, Italy}
\affiliation{Department of Physics, University of Neyshabur, P.O.Box 91136-899, Neyshabur, Iran}

\date{\today}

\begin{abstract}
Given a locally consistent set of reduced density matrices, we construct approximate density matrices which are globally consistent with the local density matrices we started from when the trial density matrix has a tree structure. We employ the cavity method of statistical physics to find the optimal density matrix representation by slowly decreasing the temperature in an annealing algorithm, or by minimizing an approximate Bethe free energy depending on the reduced density matrices and some cavity messages originated from the Bethe approximation of the entropy. We obtain the classical Bethe expression for the entropy within a naive (mean-field) approximation of the cavity messages, which is expected to work well at high temperatures. In the next order of the approximation, we obtain another expression for the Bethe entropy depending only on the diagonal elements of the reduced density matrices. In principle, we can improve the entropy approximation by considering more accurate cavity messages in the Bethe approximation of the entropy. We compare the annealing algorithm and the naive approximation of the Bethe entropy with exact and approximate numerical simulations for small and large samples of the random transverse Ising model on random regular graphs.  
\end{abstract}


\maketitle

\section{Introduction}\label{S0}
The problem of estimating the local expectation values in an interacting system is of central  
importance in classical and quantum statistical physics. This is, in general, a computationally hard problem, especially for disordered systems displaying glassy behaviors, where approximation algorithms based on the Monte Carlo sampling could be very time consuming. At least for finite-connectivity models with a locally tree-like interaction graph, the cavity method of statistical physics based on the Bethe approximation provides efficient message-passing algorithms that have proven useful in the study of random constraint satisfaction and optimization problems \cite{MP-epjb-2001,MZ-pre-2002,MPZ-science-2002,KMRSZ-pnas-2007,MM-book-2009}.\\

We can write the (Bethe) free energy for the classical Ising model on a tree using only the one-spin and the two-spin marginals of the Gibbs probability measure. On loopy graphs, this expression provides an approximate free energy, but we know how to obtain more accurate free energies by the cluster variation method and the generalized Bethe approximations taking into account the higher order correlations \cite{P-jphysa-2005,YFW-nips-2001}. Along the same lines, in this work we are going to write approximate Bethe free energies for the quantum transverse Ising model using a set of locally consistent reduced density matrices \cite{morita}. As we will see, this is not as straightforward as in the classical case, even for models on tree graphs.\\

There are various quantum cavity methods in the literature approaching the above problem from 
different perspectives \cite{CWW-prb-1992,H-prb-2007,LP-aphys-2008,LSS-prb-2008,KRSZ-prb-2008,IM-prl-2010,R-prb-2012}. Here we briefly explain the methods that are more relevant to our discussions in this study; for a review see \cite{BFKSZ-pr-2013} and references therein. The path integral quantum cavity method \cite{LSS-prb-2008,KRSZ-prb-2008} utilizes the Suzuki-Trotter transformation to map the quantum problem to a classical one and exploits the classical cavity method to estimate the local quantum expectations. The method is computationally demanding but it provides an approximate free energy density that is expected to be exact for sparse interaction graphs in the thermodynamic limit. On the other hand, the operator quantum cavity method of Refs. \cite{IM-prl-2010,DM-jstat-2010} works with one-spin cavity Hamiltonians that are determined recursively by projection from a larger cavity Hamiltonian; the latter is obtained from the neighboring one-spin cavity Hamiltonians. The method gives the local reduced density matrices in terms of the one-spin cavity Hamiltonians but it does not provide a connection between the cavity Hamiltonians and the free energy of the system. \\

In this work, we take a variational approach extending the variational quantum cavity method of Refs. \cite{R-prb-2012,RZ-prb-2012,BR-jstat-2013} to finite temperature systems; see also \cite{PH-prl-2011} and the extension of matrix product states to finite temperatures \cite{ZV-prl-2004,VGC-prl-2004,FW-prb-2005}. To this end, we first propose an approximate expression for the matrix elements of the density matrix in terms of the matrix elements of a locally consistent set of reduced density matrices. The structure (interaction graph) of the trial density matrix is chosen such that for tree interaction graphs, the locally consistent set of reduced density matrices we started from is globally consistent. Then we use the above density matrix to write down the Bethe free energy as a function of the reduced density matrices and the cavity messages that are needed to compute the entropy within the Bethe approximation. Approximating the cavity messages with a product distribution (or mean-field approximation) leads to the classical Bethe expression for the entropy \cite{morita}, which is expected to work well at high temperatures away from quantum phase transition points. We improve on this approximation by considering the two-spin marginals of the cavity messages and obtain an approximate Bethe entropy that depends only on the diagonal elements of the reduced density matrices. \\

To find the local density matrices minimizing the free energy we try two different strategies. We start from an annealing algorithm using the density matrix representation to obtain the lower-temperature reduced density matrices by the belief propagation (BP) algorithm \cite{KFL-inform-2001} relying on the Bethe approximation. As we will see, this annealing algorithm is very easy to implement but we need very accurate density matrix representations to reduce the error accumulated during the annealing process. Here we compare the results with those of the path integral quantum cavity method \cite{KRSZ-prb-2008} and exact numerical simulations of the random transverse Ising model on a random regular graph. Alternatively, we can directly minimize the approximate Bethe free energy as a function of the reduced density matrices and the cavity messages entered in the entropy approximation. This is more accurate than the annealing algorithm but computationally more expensive. Here we compare the results obtained by the mean-field approximation of the cavity messages with exact numerical simulations and those of the operator quantum cavity method \cite{IM-prl-2010,DM-jstat-2010}. \\

The paper is organized as follows. In the next section we give the main definitions and the trial density matrices we will work with in this study. In Sec. \ref{S2} we present the annealing algorithm and write the equations for updating the reduced density matrices as the temperature decreases. In Sec. \ref{S3}, we obtain an approximate expression for the Bethe free energy in terms of the reduced density matrices. In Sec. \ref{S4}, we present the optimization algorithms that we use to minimize the approximate Bethe free energy, and, finally, the concluding remarks are given in Sec. \ref{S5}. There are four appendices that give more detail of the equations and the proofs that we use in the main text.\\

\section{Definitions and the setting}\label{S1}
Consider the transverse field Ising model with Hamiltonian $H=\sum_{(ij) \in \mathcal{E}_q}H_{ij}+\sum_{i}H_i$  
where $H_{ij}\equiv -J_{ij} \sigma_i^z\sigma_j^z$, and $H_i\equiv - h_i \sigma_i^x$. The index $i=1,\dots,N$ labels the sites in the quantum interaction graph $\mathcal{E}_q$, which defines the set of interactions in the Hamiltonian. The $\sigma_i^{x,y,z}$ are the standard Pauli matrices. In the following we will work in the $\sigma^z$ representation with orthonormal basis $|\underline{\sigma} \rangle \equiv |\sigma_1 \sigma_2 \cdots \sigma_N \rangle$. 
The system in a pure state is described by the density matrix $\rho=|\Psi \rangle \langle \Psi|$ for a normalized wave function $|\Psi \rangle=\sum_{\underline{\sigma}}\psi(\underline{\sigma})|\underline{\sigma} \rangle$. And in thermal equilibrium $\rho= e^{-\beta H}/(\mathrm{Tr}e^{-\beta H})$ where $\beta=1/T$ is the inverse temperature. 

Consider a locally consistent set of reduced density matrices $\rho_i, \rho_{ij}, \dots$, where for any two reduced density matrices $\rho_A$ and $\rho_B$ that overlap on the subset of variables $A\cap B$ we have
\begin{equation}
\mathrm{Tr}_{A\setminus A\cap B} \rho_A=\mathrm{Tr}_{B\setminus A\cap B} \rho_B.
\end{equation}
The above reduced density matrices are globally consistent if they can be obtained from the same density matrix $\rho$, i.e. $\rho_A=\mathrm{Tr}_{\setminus A} \rho$ for any subset of the variables $A$.
Then we can construct approximate density matrices which, depending on the approximation, could also be globally consistent with the reduced density matrices. 
In the mean-field approximation, the density matrix is simply approximated by $\rho(\underline{\sigma};\underline{\sigma}')=\prod_i \rho_i(\sigma_i;\sigma_i')$.
In the Bethe approximation, we may write the density matrix as   
\begin{equation}\label{rho-BP-1}
\rho(\underline{\sigma};\underline{\sigma}')=\prod_i \rho_i(\sigma_i;\sigma_i')\prod_{(ij) \in \mathcal{E}}\frac{\rho_{ij}(\sigma_i,\sigma_j;\sigma_i',\sigma_j')}{\rho_i(\sigma_i;\sigma_i')\rho_j(\sigma_j;\sigma_j')}.
\end{equation}
In this study, we will assume that the quantum interaction graph $\mathcal{E}_q$ is locally tree-like, and the interaction graph $\mathcal{E}$ is equal or very close to $\mathcal{E}_q$. Note that this is an ansatz for the matrix elements of $\rho$ that is Hermitian but not necessarily positive definite. Moreover, the above density matrix can be considered as a classical model of interacting variables $(\sigma_i,\sigma_i')$ on the interaction graph $\mathcal{E}$. In appendix \ref{app-bethe-rho}, we see that when $\mathcal{E}$ is a tree and the reduced density matrices $\rho_{ij}$ and $\rho_i$ are locally consistent we have $\rho_{ij}=\mathrm{Tr}_{\setminus i,j} \rho$ and $\rho_i=\mathrm{Tr}_{\setminus i} \rho$.

More accurate density matrices can be obtained by considering interactions between a larger number of variables, for example,
\begin{equation}\label{rho-BP-2}
\rho(\underline{\sigma};\underline{\sigma}')=\prod_{(ij)\in \mathcal{E}} \rho_{ij}(\sigma_i,\sigma_j;\sigma_i',\sigma_j')\prod_{i}\frac{\rho_{i\partial i}(\sigma_i,\sigma_{\partial i};\sigma_i',\sigma_{\partial i}')}{\prod_{k\in \partial i}\rho_{ik}(\sigma_i,\sigma_k;\sigma_i',\sigma_k')},
\end{equation}
where $\sigma_{\partial i}=\{\sigma_j|j \in \partial i\}$, and $\partial i$ denotes the neighborhood set of $i$ in the interaction graph $\mathcal{E}$. In the same lines of appendix \ref{app-bethe-rho}, we can show that for tree interaction graphs $\mathcal{E}$ and locally consistent $\rho_{i\partial i}$ and $\rho_{ij}$ we have $\rho_{i\partial i}=\mathrm{Tr}_{\setminus i,\partial i} \rho$ and $\rho_{ij}=\mathrm{Tr}_{\setminus i,j} \rho$. 
 
\section{Annealing algorithm}\label{S2}
To find the density matrix that describes the equilibrium state of the system at temperature $T=1/\beta$ we start from the density matrix at infinite temperature $\rho\propto \mathbb{I}$  and slowly decrease the temperature in an annealing process. By definition of the thermal density matrix, we have   
\begin{equation}
\rho(\beta+\epsilon)=\frac{1}{Z(\beta+\epsilon)}e^{-(\beta+\epsilon)H}=\frac{Z(\beta)}{Z(\beta+\epsilon)}e^{-\epsilon H}\rho(\beta),
\end{equation}
where $Z(\beta)=\mathrm{Tr}e^{-\beta H}$. For $\epsilon \ll 1$, we can utilize the Suzuki-Trotter transformation to approximate
\begin{equation}
e^{-\epsilon H} \approx \prod_{(ij)\in \mathcal{E}_q}e^{\epsilon J_{ij} \sigma_i^z\sigma_j^z/2} \prod_ie^{\epsilon h_i \sigma_i^x}\prod_{(ij)\in \mathcal{E}_q}e^{\epsilon J_{ij} \sigma_i^z\sigma_j^z/2}.
\end{equation}
Then the lower-temperature density matrix reads
\begin{equation}
\tilde{\rho}(\underline{\sigma};\underline{\sigma}') \propto \sum_{\underline{\sigma}''}\prod_i w_{i}(\sigma_i,\sigma_i'')\prod_{(ij)\in \mathcal{E}_q}w_{ij}(\sigma_i,\sigma_j;\sigma_i'',\sigma_j'') \rho(\underline{\sigma}'';\underline{\sigma}').
\end{equation}
The weights $w_{i}$ and $w_{ij}$ come from interaction terms $h_i\sigma_i^x$ and $J_{ij} \sigma_i^z\sigma_j^z$ in the Hamiltonian, respectively,   
\begin{align}
w_{ij}(\sigma_i,\sigma_j;\sigma_i'',\sigma_j'') &\equiv e^{\epsilon J_{ij}(\sigma_i\sigma_j+\sigma_i''\sigma_j'')/2},\\
w_{i}(\sigma_i,\sigma_i'') &\equiv \cosh(\epsilon h_i)\delta_{\sigma_i,\sigma_i''}+\sinh(\epsilon h_i)\delta_{\sigma_i,-\sigma_i''}.
\end{align} 
Here, to simplify the notation, we used $\rho$ and $\tilde{\rho}$ for $\rho(\beta)$ and $\rho(\beta+\epsilon)$, respectively. 

Let us start from the mean-field approximation of the density matrix $\rho=\prod_i \rho_i$, where at each step the density matrix is updated as follows
\begin{equation}
\tilde{\rho}(\underline{\sigma};\underline{\sigma}') \propto \sum_{\underline{\sigma}''}\prod_i \Big[ w_{i}(\sigma_i,\sigma_i'')\rho_{i}(\sigma_i'';\sigma_i') \Big] \prod_{(ij)\in \mathcal{E}_q} w_{ij}(\sigma_i,\sigma_j;\sigma_i'',\sigma_j'').
\end{equation}
But this is no longer a product state and we need to project it into a mean-field state by considering only the one-spin reduced density matrices. Then, within the Bethe approximation the reduced density matrices $\tilde{\rho}_i=\mathrm{Tr}_{\setminus i}\tilde{\rho}$ are obtained by
\begin{equation}
\tilde{\rho}_i(\sigma_i;\sigma_i') \propto \sum_{\sigma_i''}\Big[ w_{i}(\sigma_i,\sigma_i'')\rho_{i}(\sigma_i'';\sigma_i') \Big] \prod_{j\in \partial_q i}\left(\sum_{\sigma_j''} w_{ij}(\sigma_i,\sigma_j;\sigma_i'',\sigma_j'') \mu_{j\to i}(\sigma_j,\sigma_j'') \right).
\end{equation}
Here, $\partial_q i$ denotes the neighborhood set of $i$ in $\mathcal{E}_q$, and the cavity marginals $\mu_{i\to j}(\sigma_i,\sigma_i'')$ are determined by the BP equations for the Gibbs measure $\tilde{\rho}(\underline{\sigma};\underline{\sigma})$ \cite{MM-book-2009}, 
\begin{equation}
\mu_{i\to j}(\sigma_i,\sigma_i'') \propto \sum_{\sigma_i''}\Big[ w_{i}(\sigma_i,\sigma_i'')\rho_{i}(\sigma_i'';\sigma_i) \Big] \prod_{k\in \partial_q i \setminus j}\left(\sum_{\sigma_k''} w_{ik}(\sigma_i,\sigma_k;\sigma_i'',\sigma_k'') \mu_{k\to i}(\sigma_k,\sigma_k'') \right).
\end{equation}
Given the $\rho_i$ and the weights $w_i,w_{ij}$, we solve the above equations by iteration starting from random initial cavity marginals $\mu_{i\to j}(\sigma_i,\sigma_i'')$, and use the cavity marginals to find the lower-temperature reduced density matrices $\tilde{\rho}_i$. 

\begin{figure}
\includegraphics[width=8cm]{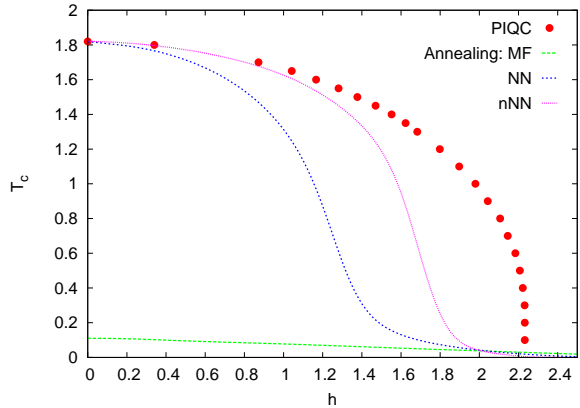}
\caption{Comparing the phase transition points in the transverse Ising model ($J_{ij}=1$ and $h_i=h$) on a random regular graph of degree $K=3$ obtained by the annealing algorithm (for different density matrix representations) and the exact solution of the path integral quantum cavity (PIQC) method \cite{KRSZ-prb-2008} in the thermodynamic limit. We take density matrices with one-spin interactions (MF), and with two-spin interactions between nearest neighbors (NN) and next-nearest neighbors (nNN) from the quantum interaction graph $\mathcal{E}_q$}\label{f1}
\end{figure}

Figure \ref{f1} shows the paramagnetic to ferromagnetic phase transition points we obtain in this way for the ferromagnetic transverse Ising model on a random regular graph. For reference, we also display the asymptotically exact results of the path integral quantum cavity method \cite{KRSZ-prb-2008}.  Unfortunately, the errors in each step of the annealing algorithm are accumulated giving rise to larger and larger errors as we decrease the temperature. The point is that in each step of the annealing process, we assume the present density matrix $\rho$ is the right density matrix at inverse temperature $\beta$, which is only correct if we worked with the most general density matrix representation. As a result, the density matrix that we obtain is not the optimal one; indeed, minimizing the free energy directly at inverse temperature $\beta$ with the same density matrix representation could result in smaller free energies. However, as the figure shows, the error is reduced by enlarging the space of the trial density matrices.  In appendix \ref{app-annealing}, we give the equations for updating some correlated density matrices with nontrivial correlation patterns as the temperature decreases. 

It is difficult to say how many interactions we need to obtain the exact behavior. At least for the ferromagnetic transverse Ising model at zero temperature, we obtain very good estimations of the ground-state properties by considering only the nearest and next-nearest neighbor interactions. That is, nearly all of the error that we observe in the annealing algorithm is the error collected all the way from infinite temperature due to the deviation of the approximated thermal state at each step from the actual one.  

\begin{figure}
\includegraphics[width=8cm]{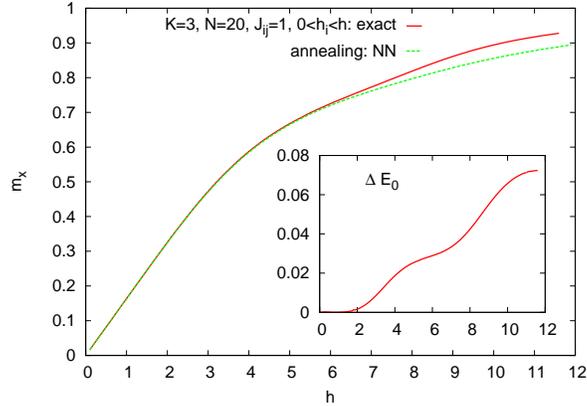}
\caption{The magnetization density $m_x$ at zero temperature in the random transverse Ising model on a random regular graph of degree $K=3$ obtained by the annealing algorithm and the exact numerical simulations for a small system of size $N=20$. The inset shows the error $\Delta E_0\equiv E_0^{ann}-E_0^{exact}$ in estimating the ground-state energy by the annealing algorithm. Here the $J_{ij}=1$ and the transverse fields $h_i$ are random numbers uniformly distributed in $[0,h]$. We take density matrices with two-spin interactions between nearest neighbors (NN) from the quantum interaction graph $\mathcal{E}_q$}\label{f2}
\end{figure}

Note that we do not have the above problem at zero temperature; the accumulated errors in the annealing algorithm that are seen for small but nonzero temperatures are not relevant at zero temperature. The fact that $\beta$ is infinity allows us to run the algorithm for a sufficiently large number of steps as in an imaginary time evolution algorithm. Then a small overlap with the ground state of the system is enough to obtain a good estimation of the ground-state properties. In Fig. \ref{f2}, we compare the algorithm predictions at zero temperature with the exact ones for the random transverse Ising model on a small random regular graph. A very smooth transition from paramagnetic to ferromagnetic phase happens around $h=9$, which is why we display the data up to $h=12$.

\section{Bethe free energy approximations}\label{S3}  
Considering the local density matrices $\rho_{i\partial i},\rho_{ij}$ and the associated density matrix $\rho$, we approximate the average energy by $\langle H \rangle=\mathrm{Tr}(\rho H)=\sum_{(ij) \in \mathcal{E}_q}\mathrm{Tr} (\rho_{ij}H_{ij})+\sum_{i}\mathrm{Tr} (\rho_{i\partial i} H_i)$. Then we utilize the replica trick to relate the entropy to a partition function in a replicated system, $S=-\mathrm{Tr}(\rho \ln \rho)=-\frac{\partial}{\partial n}\mathrm{Tr}(\rho^{n+1})|_{n=0}$. To compute the entropy we assume $n$ is an integer and consider the replicated system of interacting variables $\boldsymbol \sigma_i\equiv \{\sigma_i^0,\sigma_i^1,\dots,\sigma_i^n\}$. In the end, we will take the limit $n\to 0$. In appendix \ref{app-bethe-s}, we use the Bethe approximation to write the above entropy in terms of the reduced density matrices and the cavity messages of the Bethe approximation. 
In this way, for the entropy, we obtain
\begin{equation}
S_{Bethe}=\sum_i \frac{\partial}{\partial n} \Delta F_i|_{n=0}-\sum_{(ij)\in \mathcal{E}} \frac{\partial}{\partial n} \Delta F_{ij}|_{n=0}\equiv \sum_i \Delta s_{i}-\sum_{(ij) \in \mathcal{E}}\Delta s_{ij},
\end{equation}
where the local free energy changes $\Delta F_i$ and $\Delta F_{ij}$ are given by
\begin{align}
e^{-\Delta F_i} &= \sum_{\boldsymbol\sigma_i,\boldsymbol\sigma_{\partial i}} \rho_{i\partial i}(\boldsymbol\sigma_i,\boldsymbol\sigma_{\partial i})\prod_{j \in \partial i} \mu_{j\to i}(\boldsymbol\sigma_i,\boldsymbol\sigma_j),\\
e^{-\Delta F_{ij}} &= \sum_{\boldsymbol\sigma_i,\boldsymbol\sigma_j} \rho_{ij}(\boldsymbol\sigma_i,\boldsymbol\sigma_j) \mu_{i\to j}(\boldsymbol\sigma_i,\boldsymbol\sigma_j)\mu_{j\to i}(\boldsymbol\sigma_i,\boldsymbol\sigma_j). 
\end{align}
The $\mu_{i\to j}(\boldsymbol\sigma_i,\boldsymbol\sigma_j)$ are the cavity marginals of the replicated variables $\boldsymbol \sigma_i$ satisfying the recursive Bethe equations
\begin{equation}
\mu_{i\to j}(\boldsymbol\sigma_i,\boldsymbol\sigma_j) \propto \sum_{\boldsymbol\sigma_{\partial i\setminus j}}\frac{\rho_{i\partial i}(\boldsymbol\sigma_i,\boldsymbol\sigma_{\partial i})}{\rho_{ij}(\boldsymbol\sigma_i,\boldsymbol\sigma_j)}\prod_{k \in \partial i\setminus j }\mu_{k\to i}(\boldsymbol\sigma_i,\boldsymbol\sigma_k).
\end{equation}
We also defined the replicated density matrices
\begin{align}
\rho_{ij}(\boldsymbol \sigma_i,\boldsymbol \sigma_j) &=\prod_{t=0}^n \rho_{ij}(\sigma_i^t,\sigma_j^t;\sigma_i^{t+1},\sigma_j^{t+1}),\\
\rho_{i\partial i}(\boldsymbol\sigma_i,\boldsymbol\sigma_{\partial i}) &=\prod_{t=0}^n \rho_{i\partial i}(\sigma_i^{t},\sigma_{\partial i}^{t};\sigma_i^{t+1},\sigma_{\partial i}^{t+1}), 
\end{align}
with $\sigma_i^{n+1}=\sigma_i^{0}$ for all $i$. 

As long as the interaction graph $\mathcal{E}$ is a tree the above equations give the exact entropy for the given trial density matrix. But, to find a closed expression for the entropy we have to resort to approximations, e.g. approximating the cavity messages by a small subset of the local marginals. And working with an ansatz for the cavity messages would result in an approximate expression for the entropy. Note that using the Bethe equations for the replicated system means that we assume the replicated system is in a replica symmetric phase. All of the approximations that we will use in the following are just to simplify the equations by assuming simple structures for the joint cavity marginals of the replicas, and this is different from the well-known replica symmetry breaking approximations. 

A simple approximation for the entropy can be obtained by ignoring the correlations between the replicas, which is a mean-field approximation in the space of the replicas. More precisely, we assume $\mu_{i\to j}(\boldsymbol\sigma_i,\boldsymbol\sigma_j) \approx \prod_{t=0}^n \mu_{i\to j}^{(1)}(\sigma_i^t,\sigma_j^t)$, using only the one-spin marginals $\mu_{i\to j}^{(1)}(\sigma_i,\sigma_j)$ of the cavity messages. For tree interaction graphs $\mathcal{E}$, these marginals are simply given by $\mu_{i\to j}^{(1)}(\sigma_i,\sigma_j)|_{n=0}=1/2^2$, thanks to the marginalization properties $\rho_{ij}=\mathrm{Tr}_{\setminus i,j}\rho_{i\partial i}$. Here the entropy reads (see appendix \ref{app-bethe-s}),     
\begin{equation}\label{S-MF}
S_{Bethe}^{(1)}=-\sum_i \mathrm{Tr} (\rho_{i\partial i} \ln \rho_{i\partial i})+\sum_{(ij) \in \mathcal{E}}\mathrm{Tr} (\rho_{ij} \ln \rho_{ij}).
\end{equation}
The above entropy is indeed the classical Bethe expression for the entropy, which is expected to work well for high temperatures. As we will see, there is a temperature $T_s$ depending on the strength of the transverse fields such that for $T<T_s$, the entropy becomes negative.

Using the one- and two-spin marginals, we can approximate the cavity messages by 
\begin{align}
\mu_{i\to j}(\boldsymbol\sigma_i,\boldsymbol\sigma_j) \approx \prod_{t=0}^n \frac{\mu_{i\to j}^{(2)}(\sigma_i^t,\sigma_j^t;\sigma_i^{t+1},\sigma_j^{t+1}) }{\sqrt{\mu_{i\to j}^{(1)}(\sigma_i^t,\sigma_j^t)\mu_{i\to j}^{(1)}(\sigma_i^{t+1},\sigma_j^{t+1})} } \equiv \prod_{t=0}^n \nu_{i\to j}(\sigma_i^t,\sigma_j^t;\sigma_i^{t+1},\sigma_j^{t+1}).
\end{align}
The local marginals $\mu_{i\to j}^{(1)}(\sigma_i,\sigma_j)$ and $\mu_{i\to j}^{(2)}(\sigma_i,\sigma_j;\sigma_i',\sigma_j')$ satisfy the approximate BP equations, 
\begin{align}
\mu_{i\to j}^{(1)}(\sigma_i,\sigma_j) &\propto  \sum_{\sigma_{\partial i \setminus j}} \langle \sigma_i\sigma_j\sigma_{\partial i \setminus j}|R_{i\partial i\setminus j}^{n+1} |\sigma_i\sigma_j\sigma_{\partial i \setminus j} \rangle, \label{app-BP-1} \\ 
\mu_{i\to j}^{(2)}(\sigma_i,\sigma_j;\sigma_i',\sigma_j') &\propto  \sum_{\sigma_{\partial i \setminus j},\sigma_{\partial i \setminus j}'} \langle \sigma_i\sigma_j\sigma_{\partial i \setminus j}|R_{i\partial i\setminus j} |\sigma_i'\sigma_j'\sigma_{\partial i \setminus j}' \rangle\langle \sigma_i'\sigma_j'\sigma_{\partial i \setminus j}'|R_{i\partial i\setminus j}^n |\sigma_i\sigma_j\sigma_{\partial i \setminus j} \rangle, \label{app-BP-2}
\end{align}
where $R_{i\partial i\setminus j}$ depends on the reduced density matrices and the cavity messages,
\begin{equation}
\langle \sigma_i\sigma_j\sigma_{\partial i \setminus j}| R_{i\partial i\setminus j} |\sigma_i'\sigma_j'\sigma_{\partial i \setminus j}' \rangle \equiv \frac{\rho_{i\partial i}(\sigma_i,\sigma_{\partial i};\sigma_i',\sigma_{\partial i}')}{\rho_{ij}(\sigma_i,\sigma_j;\sigma_i',\sigma_j')} \prod_{k \in \partial i \setminus j}  \nu_{k\to i}(\sigma_i,\sigma_k;\sigma_i',\sigma_k'). 
\end{equation}
Finally for the entropy we find (see appendix \ref{app-bethe-s}),
\begin{equation}\label{S-BP}
S_{Bethe}^{(2)}=-\sum_i \mathrm{Tr} (R_{i\partial i} \ln R_{i\partial i})|_{n=0}+\sum_{(ij)\in \mathcal{E}} \mathrm{Tr} (R_{ij} \ln R_{ij})|_{n=0},
\end{equation}
where the matrix elements of $R_{i\partial i}$ and $R_{ij}$ are given by 
\begin{align}
\langle \sigma_i\sigma_{\partial i}| R_{i\partial i} |\sigma_i'\sigma_{\partial i}' \rangle &\equiv \rho_{i\partial i}(\sigma_i,\sigma_{\partial i};\sigma_i',\sigma_{\partial i}') \prod_{j \in \partial i}\nu_{j\to i}(\sigma_i,\sigma_j;\sigma_i',\sigma_j'),\\
\langle \sigma_i\sigma_j| R_{ij} |\sigma_i'\sigma_j' \rangle &\equiv \rho_{ij}(\sigma_i,\sigma_j;\sigma_i',\sigma_j') \nu_{i\to j}(\sigma_i,\sigma_j;\sigma_i',\sigma_j')\nu_{j\to i}(\sigma_i,\sigma_j;\sigma_i',\sigma_j'). 
\end{align}

Note that the entropy is computed in the limit $n\to 0$ where from the above equations we have
$\mu_{i\to j}^{(2)}(\sigma_i,\sigma_j;\sigma_i',\sigma_j') \propto \mu_{i\to j}^{(1)}(\sigma_i,\sigma_j)\delta_{\sigma_i,\sigma_i'}\delta_{\sigma_j,\sigma_j'}$. Here the matrices $R_{i\partial i}$ and $R_{ij}$ are diagonal and for the local entropy changes we obtain  
\begin{align}
\Delta s_{i}^{(2)} &=-\sum_{\sigma_i,\sigma_{\partial i}} \rho_{i\partial i}(\sigma_i,\sigma_{\partial i};\sigma_i,\sigma_{\partial i}) \ln \rho_{i\partial i}(\sigma_i,\sigma_{\partial i};\sigma_i,\sigma_{\partial i}) ,\\
\Delta s_{ij}^{(2)} &=-\sum_{\sigma_i,\sigma_j} \rho_{ij}(\sigma_i,\sigma_j;\sigma_i,\sigma_j) \ln \rho_{ij}(\sigma_i,\sigma_j;\sigma_i,\sigma_j).
\end{align}

In the same way one can improve the approximation by taking into account the higher order correlations, for example,
\begin{equation}
\mu_{i\to j}(\boldsymbol\sigma_i,\boldsymbol\sigma_j) \approx \prod_{t=0}^n\frac{\mu_{i\to j}^{(3)}(\sigma_i^t,\sigma_j^t;\sigma_i^{t+1},\sigma_j^{t+1};\sigma_i^{t+2},\sigma_j^{t+2})}{\sqrt{\mu_{i\to j}^{(2)}(\sigma_i^t,\sigma_j^t;\sigma_i^{t+1},\sigma_j^{t+1})\mu_{i\to j}^{(2)}(\sigma_i^{t+1},\sigma_j^{t+1};\sigma_i^{t+2},\sigma_j^{t+2})}}. 
\end{equation}
Note that as long as the density matrix is diagonal, we observe that for a given ansatz of the density matrix $S_{Bethe}^{(1)}=S_{Bethe}^{(2)}$. And we expect to obtain the same expression for the Bethe entropy (free energy) also in the higher orders of the approximation. 

\section{Free-energy minimization}\label{S4}
In this section, we present an optimization algorithm to estimate the optimal reduced density matrices minimizing the approximate Bethe free energy. We recall that the Bethe free energy $F_{Bethe}=\sum_{i}\langle H_i \rangle+\sum_{(ij)\in \mathcal{E}_q}\langle H_{ij} \rangle-T\left(\sum_{i}\Delta s_i-\sum_{(ij)\in \mathcal{E}}\Delta s_{ij}\right)$ is a local function of the $\rho_{i\partial i}, \rho_{ij}$ respecting the marginalization constraints, and the messages $\nu_{i\to j}$ satisfying the approximate BP equations. Let us consider the Bethe free energy as the energy function of the interacting system of variables $\rho_{i\partial i}, \rho_{ij}$ and $\nu_{ij}\equiv \{ \nu_{i\to j}, \nu_{j\to i}\}$. Then an optimization algorithm can be obtained by studying the following statistical physics problem within a higher-level Bethe approximation:
\begin{equation}
\mathcal{Z}\equiv \sum_{\{\rho_{i\partial i}\}}\sum_{\{\rho_{ij}\}}\sum_{\{\nu_{ij}\}}e^{-\beta_{opt} F_{Bethe}}\prod_{i}\prod_{j \in \partial i} \delta(\rho_{ij}-\mathrm{Tr}_{\setminus i,j} \rho_{i\partial i})\delta(\nu_{i\to j}-\hat{\nu}_{i\to j}).
\end{equation}
Here $\hat{\nu}_{i\to j}$ is a functional of the cavity messages defined by the approximate BP equations (\ref{app-BP-1}) and (\ref{app-BP-2}). And $\beta_{opt}$ is a fictitious inverse temperature to control the optimization problem. In appendix \ref{app-mp} we describe an approximate message-passing algorithm to study the above optimization problem. 

In the following we will focus on the first order of the entropy approximation given in equation \ref{S-MF}.
Here we present another message-passing algorithm, which in this case is much easier to implement than the above general algorithm.  
We will compare the numerical results with the quantum cavity method of \cite{IM-prl-2010,DM-jstat-2010} dealing with effective cavity Hamiltonians.

\subsection{Lagrangian approach}\label{41}
\begin{figure}
\includegraphics[width=8cm]{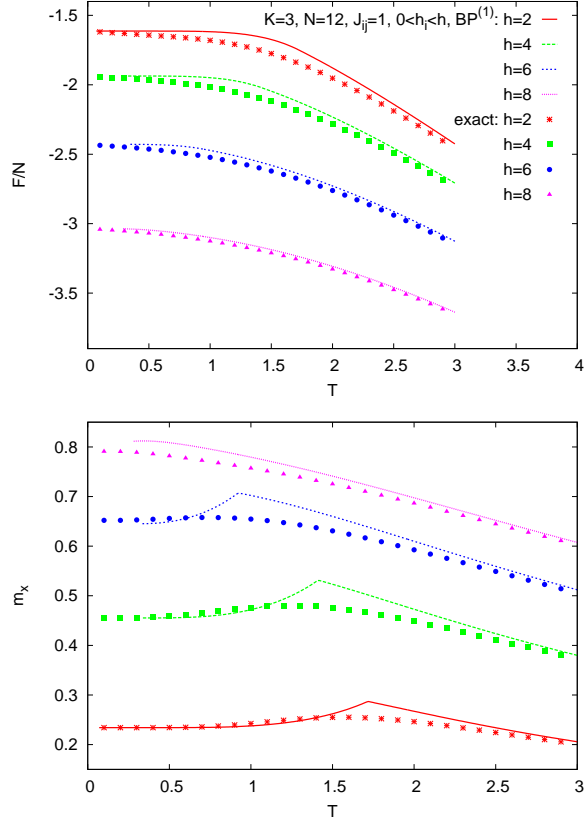}
\caption{(a) The free energy $F$ and (b) the magnetization density $m_x$ in the random transverse Ising model on a random regular graph of degree $K=3$ obtained by minimizing the approximate Bethe free energy with $S_{Bethe}^{(1)}$ (denoted by $BP^{(1)}$) and the exact numerical simulations for a small system of size $N=12$. Here the $J_{ij}=1$ and the transverse fields $h_i$ are random numbers uniformly distributed in $[0,h]$. The $BP^{(1)}$ data are shown in the region where the entropy is nonnegative.}\label{f3}
\end{figure}

An iterative algorithm to find the optimal reduced density matrices can be obtained by minimizing the following Lagrangian using Lagrange multipliers to satisfy the marginalization constraints,
\begin{equation}
\mathcal{L}\equiv F_{Bethe}^{(1)}+\sum_{i}\sum_{j\in \partial i}\left( \mathrm{Tr} (\Lambda_{ij \to i}\rho_{ij})-\mathrm{Tr} (\Lambda_{ij \to i }\rho_{i\partial i}) \right), 
\end{equation}
where $\Lambda_{ij \to i}$ is a Lagrange multiplier acting on the Hilbert space of spins $(i,j)$. We consider the mean-field approximation of the Bethe entropy $S_{Bethe}^{(1)}$ where the joint cavity marginals $\mu_{i\to j}(\boldsymbol\sigma_i,\boldsymbol\sigma_j)$ are approximated by a product distribution. Here the reduced density matrices minimizing the Lagrangian are simply given by    
\begin{align}
\rho_{i\partial i} &=\frac{1}{Z_{i\partial i}} e^{-\beta H_i+ \sum_{j \in \partial i} \Lambda_{ij\to i}},\\
\rho_{ij} &=\frac{1}{Z_{ij}} e^{\beta H_{ij}+ \Lambda_{ij\to i}+ \Lambda_{ij\to j}}.
\end{align}    
Here, for convenience, we absorb the $\beta$ into the Lagrange multipliers.
Then by the consistency of the local density matrices we obtain
\begin{align}
\Lambda_{ij  \to j}=-\beta H_{ij}-\Lambda_{ij \to i}+\ln \left( \frac{ Z_{ij}}{Z_{i\partial i}}\mathrm{Tr}_{\setminus i,j} e^{-\beta H_i+  \sum_{k \in \partial i} \Lambda_{ik \to i}} \right).
\end{align}
These equations can be solved by iteration starting from Hermitian $\Lambda_{ij \to i}$. This is enough to ensure that the resulting reduced density matrices are Hermitian and positive semidefinite. Figure \ref{f3} displays the results obtained in this way along with the exact solution for a small system of random transverse Ising model on a random regular graph. As expected, the predictions are in good agreement with the exact ones for high temperatures, but the difference is larger close to the phase transition points and the entropy becomes negative for small temperatures $T<T_s$, where $T_s$ is an increasing function of the transverse fields.

\begin{figure}
\includegraphics[width=8cm]{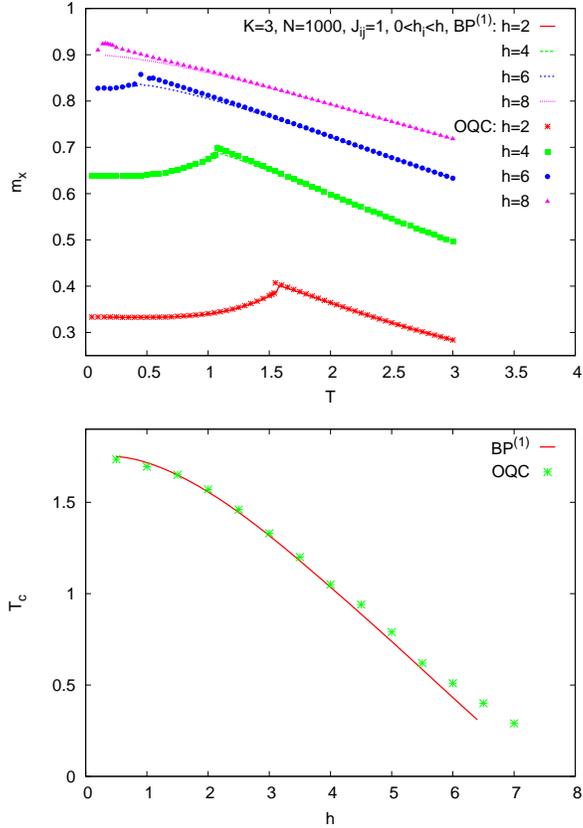}
\caption{Comparing (a) the magnetization density $m_x$ and (b) the phase transition points $T_c$ in the random transverse Ising model on a random regular graph of degree $K=3$ obtained by minimizing the approximate Bethe free energy with $S_{Bethe}^{(1)}$ (denoted by $BP^{(1)}$) and the operator quantum cavity method (OQC) of Refs. \cite{IM-prl-2010,DM-jstat-2010} for a system of size $N=1000$.  Here the $J_{ij}=1$ and the transverse fields $h_i$ are random numbers uniformly distributed in $[0,h]$. The $BP^{(1)}$ data are shown in the region where the entropy is nonnegative.}\label{f4}
\end{figure}
 
Let us compare the above equations with the ones obtained by the quantum cavity method of Refs. \cite{IM-prl-2010,DM-jstat-2010}, where the reduced density matrices are given by  
\begin{align}
\rho_{i\partial i} &=\frac{1}{Z_{i\partial i}} e^{-\beta \tilde{H}_{i\partial i}},\hskip1cm \tilde{H}_{i\partial i}= H_i+\sum_{j \in \partial i}( H_{ij}+ \tilde{H}_{j\to i}),\\
\rho_{ij} &=\frac{1}{Z_{ij}} e^{-\beta \tilde{H}_{ij}}, \hskip1cm  \tilde{H}_{ij}=H_{ij}+ \tilde{H}_{i\to j}+ \tilde{H}_{j\to i}.
\end{align} 
As before, $H_i=-h_i\sigma_i^x$ and $H_{ij}=-J_{ij}\sigma_i^z\sigma_j^z$.   
The cavity Hamiltonians are determined as follows \cite{DM-jstat-2010}: Using the one-spin Hamiltonians $\tilde{H}_{k \to i}$, we first write the cavity Hamiltonian
\begin{align}
\tilde{H}_{i\partial i \to j}=-h_i \sigma_i^x+\sum_{k\in \partial i \setminus j}(-J_{ik}\sigma_i^z\sigma_k^z+ \tilde{H}_{k\to i}).
\end{align}    
Then we obtain $\tilde{H}_{i\to j}=-h_i \sigma_i^x-g_{i\to j}\sigma_i^z$ by finding the
$g_{i\to j}$ such that $\mathrm{Tr}(\rho_{i\to j}\sigma_i^z)=\mathrm{Tr}(\rho_{i\partial i\to j}\sigma_i^z)$, where $\rho_{i\to j} \propto e^{-\beta \tilde{H}_{i\to j}}$ and $\rho_{i\partial i\to j} \propto e^{-\beta \tilde{H}_{i\partial i \to j}}$.
In Fig. \ref{f4}, we compare the numerical results obtained by the above two algorithms.
As long as the approximate Bethe entropy is positive the two algorithms give very close estimations of the local quantum expectations and the phase transition points. However, the naive approximation of the Bethe entropy results in negative entropies at low temperatures. Equivalently, we observe that the two definitions of the free energy
\begin{align} 
F_1 &=-\sum_i \frac{1}{\beta}\ln \mathrm{Tr}  e^{-\beta \tilde{H}_{i\partial i}}+\sum_{(ij) \in \mathcal{E}_q} \frac{1}{\beta}\ln \mathrm{Tr}  e^{-\beta \tilde{H}_{ij}},\\
F_2 &=\sum_i \mathrm{Tr} (\rho_{i\partial i}H_i)+\sum_{(ij) \in \mathcal{E}_q} \mathrm{Tr}(\rho_{ij}H_{ij})-T\left (-\sum_i \mathrm{Tr} (\rho_{i\partial i}\ln \rho_{i\partial i})+\sum_{(ij) \in \mathcal{E}_q} \mathrm{Tr}(\rho_{ij}\ln \rho_{ij}) \right),
\end{align}     
in the latter algorithm are not always consistent, resulting in different free energy values. However, in numerical simulations, we observe that at least for small problem sizes, the first expression for the free energy is closer to the exact free energy.

\section{Conclusion}\label{S5}
The main question we started from was to find an approximate density matrix and free energy for a quantum system given a set of locally consistent reduced density matrices. We know how to do this by the (generalized) Bethe approximation in a classical system and our goal was to extend that construction to quantum systems. Then, the expression for free energy can be considered as a function of the reduced density matrices to compute the physical density matrices minimizing the approximate free energy. Note that as for the Bethe approximation in classical systems, the free energies we obtain are not necessarily an upper bound for the exact free energy.  

We started from an appropriate ansatz for the density matrix and used the replica trick to relate the computation of the quantum entropy to the computation of a partition function in a replicated system. We computed the replicated partition function within the Bethe approximation. Here a product (mean-field) ansatz for the cavity messages (i.e., independent replicas) resulted in the classical Bethe expression for the entropy. This clarifies the nature of the approximation we make when we replace the quantum entropy with the classical Bethe entropy. 

The leading order of the approximation with independent replicas works well for high temperatures, but results in negative entropies for very small temperatures. At this level the algorithm is easy to implement and faster than the operator quantum cavity method we used for comparison in figure \ref{f4}. The latter algorithm is, of course, more accurate for low temperatures but, as we mentioned in the previous section, it does not provide a consistent free energy approximation. Perhaps the path integral quantum cavity method is more complete in this sense but at the same time it is computationally more expensive.   

The free energy approximations can be systematically improved by considering more accurate density matrices and approximations for the cavity messages in the Bethe approximation of the entropy. In the second order of the approximation, we considered the two-spin correlations between the replicas and obtained another expression for the entropy involving only the diagonal elements of the reduced density matrices. We will further investigate this entropy and the higher orders of the approximation in future works. It would also be interesting to see how the method can be generalized to study fermionic systems at finite temperatures.

\acknowledgments
We are grateful to G. Semerjian for reading the manuscript and helpful comments. We would like to thank F. Zamponi for providing the PIQC data displayed in Fig. \ref{f1}. A.R. acknowledges support from ERC Grant No. OPTINF  267915.

\appendix

\section{Locally and globally consistent reduced density matrices}\label{app-bethe-rho}
Consider the following ansatz for the density matrix  
\begin{equation}
\rho(\underline{\sigma};\underline{\sigma}')=\prod_i \rho_i(\sigma_i;\sigma_i')\prod_{(ij) \in \mathcal{E}}\frac{\rho_{ij}(\sigma_i,\sigma_j;\sigma_i',\sigma_j')}{\rho_i(\sigma_i;\sigma_i')\rho_j(\sigma_j;\sigma_j')}.
\end{equation}
Here we prove that when $\mathcal{E}$ is a tree and the reduced density matrices are locally consistent we have $\rho_{ij}=\mathrm{Tr}_{\setminus i,j} \rho$ and $\rho_i=\mathrm{Tr}_{\setminus i} \rho$.

Let us start from computing $\mathrm{Tr} \rho$ to show that $\rho$ is trace normalized when $\rho_i=\mathrm{Tr}_j\rho_{ij}$ and $\mathrm{Tr}\rho_{i}=\mathrm{Tr}\rho_{ij}=1$.
Expanding the trace we have
\begin{equation}
Z=\mathrm{Tr}\rho= \sum_{\underline{\sigma}}\prod_{i} \rho_i(\sigma_i;\sigma_i)\prod_{(ij) \in \mathcal{E}}\frac{\rho_{ij}(\sigma_i,\sigma_j;\sigma_i,\sigma_j)}{\rho_i(\sigma_i;\sigma_i)\rho_j(\sigma_j;\sigma_j)}.     
\end{equation}
For tree structures we can write the above sum as 
\begin{equation}
Z= \sum_{\sigma_i}\rho_i(\sigma_i;\sigma_i)\prod_{j \in \partial i }\left(\sum_{\sigma_j} \frac{\rho_{ij}(\sigma_i,\sigma_j;\sigma_i,\sigma_j)}{\rho_i(\sigma_i;\sigma_i)\rho_j(\sigma_j;\sigma_j)} Z_{j\to i}(\sigma_j)\right).   
\end{equation}
Here the $Z_{j\to i}(\sigma_j)$ are the cavity partition functions computed in the absence of site $i$, where the partition function reads $\prod_{j\in \partial i}(\sum_{\sigma_j}Z_{j\to i}(\sigma_j))$. The cavity partition functions are computed recursively by the Bethe equations \cite{MM-book-2009},
\begin{equation}
Z_{i\to j}(\sigma_i)= \rho_i(\sigma_i;\sigma_i)\prod_{k \in \partial i \setminus j}\left(\sum_{\sigma_k} \frac{\rho_{ik}(\sigma_i,\sigma_k;\sigma_i,\sigma_k)}{\rho_i(\sigma_i;\sigma_i)\rho_k(\sigma_k;\sigma_k)} Z_{k\to i}(\sigma_k)\right),     
\end{equation} 
Note that for the leaves we have $Z_{i\to j}(\sigma_i)= \rho_i(\sigma_i;\sigma_i)$ and from the marginalization relations $\rho_{i}= \mathrm{Tr}_j \rho_{ij}$  we find $Z_{i\to j}(\sigma_i)= \rho_i(\sigma_i;\sigma_i)$ for all of the cavity partition functions. Therefore, we obtain $Z=\sum_{\sigma_i}\rho_i(\sigma_i;\sigma_i)=1$.

To compute the one-spin reduced density matrices, we use again the recursive equations to write
\begin{equation}
\langle \sigma_i |\mathrm{Tr}_{\setminus i} \rho |\sigma_i' \rangle= \rho_i(\sigma_i;\sigma_i') \prod_{j \in \partial i }\left(\sum_{\sigma_j} \frac{\rho_{ij}(\sigma_i,\sigma_j;\sigma_i',\sigma_j)}{\rho_i(\sigma_i;\sigma_i')\rho_j(\sigma_j;\sigma_j)} Z_{j\to i}(\sigma_j)\right).    
\end{equation}
But $Z_{j\to i}(\sigma_j)=\rho_j(\sigma_j;\sigma_j)$ which, along with the marginalization relations, give $\mathrm{Tr}_{\setminus i} \rho= \rho_i$. Similarly, one can prove that $\mathrm{Tr}_{\setminus i,j} \rho= \rho_{ij}$ thanks to the tree interaction graph $\mathcal{E}$ and the consistency of the local density matrices $\rho_i$ and $\rho_{ij}$. 

One can easily extend the above arguments to more general density matrices with higher order interactions,     
\begin{equation}
\rho(\underline{\sigma};\underline{\sigma}')=\prod_i \rho_i(\sigma_i;\sigma_i')\prod_{a}\frac{\rho_{a}(\sigma_{\partial a};\sigma_{\partial a}')}{\prod_{i \in \partial a}\rho_i(\sigma_i;\sigma_i')},
\end{equation}
as long as the bipartite graph representing the dependency of the interactions to the variables is a tree. Here, $\sigma_{\partial a}\equiv \{\sigma_i| i \in \partial a\}$ and $\partial a$ defines the set of variables in interaction $a$.

\section{Computing the reduced density matrices in the annealing algorithm}\label{app-annealing}
Consider the following ansatz for the density matrix
\begin{equation}
\rho(\underline{\sigma};\underline{\sigma}')=\prod_i \rho_i(\sigma_i;\sigma_i')\prod_{(ij) \in \mathcal{E}}\frac{\rho_{ij}(\sigma_i,\sigma_j;\sigma_i',\sigma_j')}{\rho_i(\sigma_i;\sigma_i')\rho_j(\sigma_j;\sigma_j')}.
\end{equation}
In each step of the annealing process, we need to compute the local reduced density matrices given the updated density matrix, 
\begin{equation}
\tilde{\rho}(\underline{\sigma};\underline{\sigma}') \propto \sum_{\underline{\sigma}''}\prod_i \Big[ w_{i}(\sigma_i,\sigma_i'')\rho_{i}(\sigma_i'';\sigma_i') \Big] \prod_{(ij)\in \mathcal{E}} \Big[ w_{ij}(\sigma_i,\sigma_j;\sigma_i'',\sigma_j'')\frac{\rho_{ij}(\sigma_i'',\sigma_j'';\sigma_i',\sigma_j')}{\rho_{i}(\sigma_i'';\sigma_i')\rho_{i}(\sigma_j'';\sigma_j')} \Big],
\end{equation}
where, for simplicity, we assumed $\mathcal{E}=\mathcal{E}_q$.
The local density matrix $\tilde{\rho}_{ij}=\mathrm{Tr}_{\setminus i,j}\tilde{\rho}$ is obtained from the above expression after summing over the $\sigma_k'=\sigma_k$ for $k\neq i,j$. For tree interaction graphs $\mathcal{E}$, this sum can be computed by considering the cavity messages that the boundary variables $\partial (ij)$ receive from the other parts of the system in addition to the local weights,      
\begin{multline}
\tilde{\rho}_{ij}(\sigma_i,\sigma_j;\sigma_i',\sigma_j') \propto \sum_{\sigma_i'',\sigma_j''} w_{ij}(\sigma_i,\sigma_j;\sigma_i'',\sigma_j'')\rho_{ij}(\sigma_i'',\sigma_j'';\sigma_i',\sigma_j') \\ \times w_{i}(\sigma_i,\sigma_i'')\prod_{k\in \partial i \setminus j}\Big[ \sum_{\sigma_k,\sigma_k''}
w_{ik}(\sigma_i,\sigma_k;\sigma_i'',\sigma_k'')\frac{\rho_{ik}(\sigma_i'',\sigma_k'';\sigma_i',\sigma_k)}{\rho_{i}(\sigma_i'';\sigma_i')\rho_{k}(\sigma_k'';\sigma_k)}\mu_{k\to i}(\sigma_k;\sigma_k'')\Big]\\ \times
w_{j}(\sigma_j,\sigma_j'')\prod_{k\in \partial j \setminus i}\Big[ \sum_{\sigma_k,\sigma_k''}
w_{jk}(\sigma_j,\sigma_k;\sigma_j'',\sigma_k'')\frac{\rho_{jk}(\sigma_j'',\sigma_k'';\sigma_j',\sigma_k)}{\rho_{j}(\sigma_j'';\sigma_j')\rho_{k}(\sigma_k'';\sigma_k)}\mu_{k\to j}(\sigma_k;\sigma_k'')\Big].
\end{multline}
Here the cavity messages $\mu_{i\to j}$ are determined recursively by the Bethe equations \cite{MM-book-2009}, 
\begin{multline}
\mu_{i\to j}(\sigma_i;\sigma_i'') \propto w_{i}(\sigma_i;\sigma_i'')\rho_{i}(\sigma_i'';\sigma_i) \\ \times \prod_{k\in \partial i \setminus j}\Big[\sum_{\sigma_k,\sigma_k''} w_{ik}(\sigma_i,\sigma_k;\sigma_i'',\sigma_k'')\frac{\rho_{ik}(\sigma_i'',\sigma_k'';\sigma_i,\sigma_k)}{\rho_{i}(\sigma_i'';\sigma_i)\rho_{k}(\sigma_k'';\sigma_k)} \mu_{k\to i}(\sigma_k;\sigma_k'')\Big].
\end{multline}

\section{Derivation of the Bethe entropy from the Bethe density matrices}\label{app-bethe-s}
Consider the following ansatz for the density matrix
\begin{equation}
\rho(\underline{\sigma};\underline{\sigma}')=\prod_{(ij)\in \mathcal{E}} \rho_{ij}(\sigma_i,\sigma_j;\sigma_i',\sigma_j')\prod_{i}\frac{\rho_{i\partial i}(\sigma_i,\sigma_{\partial i};\sigma_i',\sigma_{\partial i}')}{\prod_{k\in \partial i}\rho_{ik}(\sigma_i,\sigma_k;\sigma_i',\sigma_k')},
\end{equation}
and write the entropy as
\begin{equation}
S=- \mathrm{Tr} (\rho \ln \rho)=-\frac{\partial}{\partial n}\mathrm{Tr} \rho^{n+1}|_{n=0}.
\end{equation}
We rewrite $Z_{n+1}\equiv \mathrm{Tr} \rho^{n+1}= \sum_{\underline{\sigma}^0,\underline{\sigma}^1,\dots,\underline{\sigma}^n} \prod_{t=0}^n\rho(\underline{\sigma}^t;\underline{\sigma}^{t+1})$ with $\underline{\sigma}^{n+1}=\underline{\sigma}^{0}$ as
\begin{equation}
Z_{n+1}= \sum_{\boldsymbol \sigma_1,\boldsymbol \sigma_2,\dots,\boldsymbol \sigma_N}\prod_{(ij)\in \mathcal{E}} \rho_{ij}(\boldsymbol \sigma_i,\boldsymbol \sigma_j)\prod_{i}\frac{\rho_{i\partial i}(\boldsymbol\sigma_i,\boldsymbol\sigma_{\partial i})}{\prod_{k\in \partial i}\rho_{ik}(\boldsymbol\sigma_i,\boldsymbol\sigma_k)},
\end{equation}
where $\boldsymbol \sigma_i\equiv \{\sigma_i^0,\sigma_i^1,\dots,\sigma_i^n\}$, $\rho_{ij}(\boldsymbol \sigma_i,\boldsymbol \sigma_j)=\prod_{t=0}^n \rho_{ij}(\sigma_i^t,\sigma_j^t;\sigma_i^{t+1},\sigma_j^{t+1})$, and $\rho_{i\partial i}(\boldsymbol\sigma_i,\boldsymbol\sigma_{\partial i})=\prod_{t=0}^n \rho_{i\partial i}(\sigma_i^{t},\sigma_{\partial i}^{t};\sigma_i^{t+1},\sigma_{\partial i}^{t+1})$. Now, using the recursive Bethe equations, we have
\begin{align}
Z_{n+1}=\sum_{\boldsymbol\sigma_i,\boldsymbol\sigma_{\partial i}}\rho_{i\partial i}(\boldsymbol\sigma_i,\boldsymbol\sigma_{\partial i})\prod_{j\in \partial i}Z_{j\to i}(\boldsymbol\sigma_i,\boldsymbol\sigma_j),
\end{align}
where $Z_{i\to j}(\boldsymbol\sigma_i,\boldsymbol\sigma_j)$ is the cavity partition function given $\boldsymbol\sigma_i$ and $\boldsymbol\sigma_j$,
\begin{equation}
Z_{i\to j}(\boldsymbol\sigma_i,\boldsymbol\sigma_j)=\sum_{\boldsymbol\sigma_{\partial i\setminus j}}\frac{\rho_{i\partial i}(\boldsymbol\sigma_i,\boldsymbol\sigma_{\partial i})}{\rho_{ij}(\boldsymbol\sigma_i,\boldsymbol\sigma_j)}\prod_{k \in \partial i\setminus j }Z_{k\to i}(\boldsymbol\sigma_i,\boldsymbol\sigma_k).
\end{equation}
The belief propagation (BP) equations are equations for the normalized cavity partitions $\mu_{i\to j}(\boldsymbol\sigma_i,\boldsymbol\sigma_j)=Z_{i\to j}(\boldsymbol\sigma_i,\boldsymbol\sigma_j)/(\sum_{\boldsymbol\sigma_i',\boldsymbol\sigma_j'}Z_{i\to j}(\boldsymbol\sigma_i',\boldsymbol\sigma_j'))$,
\begin{equation}
\mu_{i\to j}(\boldsymbol\sigma_i,\boldsymbol\sigma_j) \propto \sum_{\boldsymbol\sigma_{\partial i\setminus j}}\frac{\rho_{i\partial i}(\boldsymbol\sigma_i,\boldsymbol\sigma_{\partial i})}{\rho_{ij}(\boldsymbol\sigma_i,\boldsymbol\sigma_j)}\prod_{k \in \partial i\setminus j }\mu_{k\to i}(\boldsymbol\sigma_i,\boldsymbol\sigma_k).
\end{equation}
Then the replicated partition function reads  \cite{MM-book-2009},
\begin{equation}
Z_{n+1}=\prod_i e^{-\Delta F_i}\prod_{(ij)\in \mathcal{E}} e^{\Delta F_{ij}}, 
\end{equation}
where $\Delta F_i$ is the free energy change by adding variable $\boldsymbol \sigma_i$ and the interactions involving the variable,
\begin{equation}
e^{-\Delta F_i}= \sum_{\boldsymbol\sigma_i,\boldsymbol\sigma_{\partial i}} \rho_{i\partial i}(\boldsymbol\sigma_i,\boldsymbol\sigma_{\partial i})\prod_{j \in \partial i} \mu_{j\to i}(\boldsymbol\sigma_i,\boldsymbol\sigma_j). 
\end{equation}
And $\Delta F_{ij}$ is the free energy change by adding the interaction $\rho_{ij}(\boldsymbol \sigma_i,\boldsymbol \sigma_j)$,
\begin{equation}
e^{-\Delta F_{ij}}= \sum_{\boldsymbol\sigma_i,\boldsymbol\sigma_j} \rho_{ij}(\boldsymbol\sigma_i,\boldsymbol\sigma_j) \mu_{i\to j}(\boldsymbol\sigma_i,\boldsymbol\sigma_j)\mu_{j\to i}(\boldsymbol\sigma_i,\boldsymbol\sigma_j). 
\end{equation}

In this way, for the entropy, we find
\begin{equation}
S_{Bethe}=\sum_i \frac{\partial}{\partial n} \Delta F_i|_{n=0}-\sum_{(ij)\in \mathcal{E}} \frac{\partial}{\partial n} \Delta F_{ij}|_{n=0},
\end{equation}
where we used the fact that $Z_{n+1}|_{n=0}=\mathrm{Tr} \rho=1$. 

\subsection{Approximating the cavity messages}
To compute the free energy changes, we need to resort to some reasonable approximations for the cavity message $\mu_{i\to j}(\boldsymbol\sigma_i,\boldsymbol\sigma_j)$ even when the interaction graph $\mathcal{E}$ is a tree. Note that for the messages from the leaves we have
$\mu_{i\to j}(\boldsymbol\sigma_i,\boldsymbol\sigma_j)=1/2^{2n+2}$. In fact, for $n=0$ where $\sum_{\boldsymbol\sigma_{\partial i\setminus j}} \rho_{i\partial i}(\boldsymbol\sigma_i,\boldsymbol\sigma_{\partial i}) = \rho_{ij}(\boldsymbol\sigma_i,\boldsymbol\sigma_j)$ holds we obtain $Z_{i\to j}(\boldsymbol\sigma_i,\boldsymbol\sigma_j)=1$ for all the cavity partition functions. For $n>0$, we can no longer rely on the marginalization property and the cavity partition functions could be different from one.

Let us continue by approximating all of the cavity messages by $\mu_{i\to j}(\boldsymbol\sigma_i,\boldsymbol\sigma_j) \propto 1$ as it happens for $n=0$.
Within this approximation, we find
\begin{align}
\frac{\partial}{\partial n} \Delta F_i|_{n=0} &=-\frac{\partial}{\partial n} \mathrm{Tr} \rho_{i\partial i}^{n+1}|_{n=0}+\sum_{j \in \partial i} \frac{\partial}{\partial n} 2^{2n+2}|_{n=0},\\
\frac{\partial}{\partial n} \Delta F_{ij}|_{n=0} &=-\frac{\partial}{\partial n} \mathrm{Tr} \rho_{ij}^{n+1}|_{n=0}+\frac{\partial}{\partial n} 2^{2n+2}|_{n=0}+ \frac{\partial}{\partial n} 2^{2n+2}|_{n=0},
\end{align}
where we used the fact that $\mathrm{Tr} \rho_{i\partial i}^{n+1}|_{n=0}=1$ and $\mathrm{Tr} \rho_{ij}^{n+1}|_{n=0}=1$. Moreover, $\frac{\partial}{\partial n} \mathrm{Tr} \rho_{i\partial i}^{n+1}|_{n=0}=\mathrm{Tr} (\rho_{i\partial i} \ln \rho_{i\partial i})$ and $\frac{\partial}{\partial n} \mathrm{Tr} \rho_{ij}^{n+1}|_{n=0}=\mathrm{Tr} (\rho_{ij} \ln \rho_{ij})$, resulting in 
\begin{equation}
S_{Bethe}^{(1)}=-\sum_i \mathrm{Tr} (\rho_{i\partial i} \ln \rho_{i\partial i})+\sum_{(ij)\in \mathcal{E}} \mathrm{Tr} (\rho_{ij} \ln \rho_{ij}).
\end{equation}
As we will see, the above entropy can be obtained by a mean-field approximation of the cavity messages when the interaction graph $\mathcal{E}$ is a tree.  

\subsubsection{Bethe approximation of the cavity messages}
More systematic approximations for the entropy can be obtained by writing the cavity messages
in terms of the associated local marginals. For example, using the two-spin marginals we approximate the cavity messages by
\begin{equation}
\mu_{i\to j}(\boldsymbol\sigma_i,\boldsymbol\sigma_j)= \prod_{t=0}^n \nu_{i\to j}(\sigma_i^t,\sigma_j^t;\sigma_i^{t+1},\sigma_j^{t+1}),
\end{equation}
where
\begin{equation}
\nu_{i\to j}(\sigma_i,\sigma_j;\sigma_i',\sigma_j')=\frac{\mu_{i\to j}^{(2)}(\sigma_i,\sigma_j;\sigma_i',\sigma_j')}{\sqrt{\mu_{i\to j}^{(1)}(\sigma_i,\sigma_j)\mu_{i\to j}^{(1)}(\sigma_i',\sigma_j')}}.
\end{equation}
Now taking the BP equations for the cavity messages $\mu_{i\to j}(\boldsymbol\sigma_i,\boldsymbol\sigma_j)$ and summing over $\{(\sigma_i^t,\sigma_j^t)|t=1,\dots,n\}$ we obtain
\begin{equation}\label{nu}
\langle \sigma_i^0\sigma_j^0| \nu_{i\to j}^{n+1} |\sigma_i^0\sigma_j^0 \rangle  \propto  \sum_{\sigma_{\partial i \setminus j}^0}\langle \sigma_i^0\sigma_j^0\sigma_{\partial i \setminus j}^0| R_{i\partial i\setminus j}^{n+1} |\sigma_i^0\sigma_j^0\sigma_{\partial i \setminus j}^0 \rangle,
\end{equation}
where $\langle \sigma_i\sigma_j| \nu_{i\to j} |\sigma_i'\sigma_j' \rangle \equiv \nu_{i\to j}(\sigma_i,\sigma_j;\sigma_i',\sigma_j') $ and
\begin{equation}\label{R-BP}
\langle \sigma_i\sigma_j\sigma_{\partial i \setminus j}| R_{i\partial i\setminus j} |\sigma_i'\sigma_j'\sigma_{\partial i \setminus j}' \rangle \equiv \frac{\rho_{i\partial i}(\sigma_i,\sigma_{\partial i};\sigma_i',\sigma_{\partial i}')}{\rho_{ij}(\sigma_i,\sigma_j;\sigma_i',\sigma_j')} \prod_{k \in \partial i \setminus j}  \nu_{k\to i}(\sigma_i,\sigma_k;\sigma_i',\sigma_k'). 
\end{equation}

Then, using the translational symmetry, the one-spin marginals $\mu_{i\to j}^{(1)}(\sigma_i,\sigma_j)$ read
\begin{equation}\label{nu-1}
\mu_{i\to j}^{(1)}(\sigma_i,\sigma_j)  \propto  \sum_{\sigma_{\partial i \setminus j}} \langle \sigma_i\sigma_j\sigma_{\partial i \setminus j}|R_{i\partial i\setminus j}^{n+1} |\sigma_i\sigma_j\sigma_{\partial i \setminus j} \rangle.
\end{equation}
For the two-spin marginals $\mu_{i\to j}^{(2)}(\sigma_i,\sigma_j;\sigma_i',\sigma_j')$ we obtain
\begin{equation}\label{nu-2}
\mu_{i\to j}^{(2)}(\sigma_i,\sigma_j;\sigma_i',\sigma_j') \propto  \sum_{\sigma_{\partial i \setminus j},\sigma_{\partial i \setminus j}'} \langle \sigma_i\sigma_j\sigma_{\partial i \setminus j}|R_{i\partial i\setminus j} |\sigma_i'\sigma_j'\sigma_{\partial i \setminus j}' \rangle\langle \sigma_i'\sigma_j'\sigma_{\partial i \setminus j}'|R_{i\partial i\setminus j}^n |\sigma_i\sigma_j\sigma_{\partial i \setminus j} \rangle.
\end{equation}

Finally, given the $\nu_{i\to j}(\sigma_i,\sigma_j;\sigma_i',\sigma_j')$  we end up with the following Bethe entropy
\begin{equation}
S_{Bethe}^{(2)}=-\sum_i \mathrm{Tr} (R_{i\partial i} \ln R_{i\partial i})|_{n=0}+\sum_{(ij)\in \mathcal{E}} \mathrm{Tr} (R_{ij} \ln R_{ij})|_{n=0},
\end{equation}
where
\begin{equation}
\langle \sigma_i\sigma_{\partial i}| R_{i\partial i} |\sigma_i'\sigma_{\partial i}' \rangle \equiv \rho_{i\partial i}(\sigma_i,\sigma_{\partial i};\sigma_i',\sigma_{\partial i}') \prod_{j \in \partial i}\nu_{j\to i}(\sigma_i,\sigma_j;\sigma_i',\sigma_j'),
\end{equation}
and
\begin{equation}
\langle \sigma_i\sigma_j| R_{ij} |\sigma_i'\sigma_j' \rangle \equiv \rho_{ij}(\sigma_i,\sigma_j;\sigma_i',\sigma_j') \nu_{i\to j}(\sigma_i,\sigma_j;\sigma_i',\sigma_j')\nu_{j\to i}(\sigma_i,\sigma_j;\sigma_i',\sigma_j'). 
\end{equation}
Note that in computing the entropy, we ignored $-\sum_i \mathrm{Tr} (\frac{\partial}{\partial n} R_{i\partial i})|_{n=0}+\sum_{(ij)\in \mathcal{E}} \mathrm{Tr} (\frac{\partial}{\partial n}R_{ij})|_{n=0}$ as the entropy is stationary with respect to the changes in the cavity messages. 

In the same way, one can improve the approximation by taking into account the higher order correlations, for example, assuming
\begin{equation}
\mu_{i\to j}(\boldsymbol\sigma_i,\boldsymbol\sigma_j)= \prod_{t=0}^n\nu_{i\to j}(\sigma_i^{t-1},\sigma_j^{t-1};\sigma_i^{t},\sigma_j^{t};\sigma_i^{t+1},\sigma_j^{t+1}), 
\end{equation}
with
\begin{equation}
\nu_{i\to j}(\sigma_i',\sigma_j';\sigma_i,\sigma_j;\sigma_i'',\sigma_j'')=\frac{\mu_{i\to j}^{(3)}(\sigma_i',\sigma_j';\sigma_i,\sigma_j;\sigma_i'',\sigma_j'')}{\sqrt{\mu_{i\to j}^{(2)}(\sigma_i',\sigma_j';\sigma_i,\sigma_j)\mu_{i\to j}^{(2)}(\sigma_i,\sigma_j;\sigma_i'',\sigma_j'')}}.
\end{equation}

\subsubsection{The mean-field approximation of the cavity messages}   
Consider the mean-field approximation of the cavity messages $\mu_{i\to j}(\boldsymbol\sigma_i,\boldsymbol\sigma_j)=\prod_{t=0}^n \mu_{i\to j}^{(1)}(\sigma_i^t,\sigma_j^t)$ where
$\nu_{i\to j}(\sigma_i,\sigma_j;\sigma_i',\sigma_j')=\sqrt{\mu_{i\to j}^{(1)}(\sigma_i,\sigma_j)\mu_{i\to j}^{(1)}(\sigma_i',\sigma_j')}$. To compute the entropy, we need $\mu_{i\to j}^{(1)}(\sigma_i,\sigma_j)|_{n=0}$ which according to Eqs. \ref{R-BP} and \ref{nu-1} reads
\begin{equation}
\mu_{i\to j}^{(1)}(\sigma_i,\sigma_j)  \propto  \sum_{\sigma_{\partial i \setminus j}}  \frac{\rho_{i\partial i}(\sigma_i,\sigma_{\partial i};\sigma_i,\sigma_{\partial i})}{\rho_{ij}(\sigma_i,\sigma_j;\sigma_i,\sigma_j)} \prod_{k \in \partial i \setminus j}  \mu_{k\to i}^{(1)}(\sigma_i,\sigma_k). 
\end{equation}
Suppose the interaction graph $\mathcal{E}$ is a tree. Then, for the messages from the leaves we have $\mu_{i\to j}^{(1)}(\sigma_i,\sigma_j)=1/2^2$ and using the marginalization property $\rho_{ij}=\mathrm{Tr}_{\setminus ij} \rho_{i\partial i}$ we find that indeed $\mu_{i\to j}^{(1)}(\sigma_i,\sigma_j)=1/2^2$ holds for all of the messages. Consequently, we recover the classical expression for the Bethe entropy $S_{Bethe}^{(1)}$ as described at the beginning of this section.

\section{A message-passing algorithm for minimizing the approximate Bethe free energy}\label{app-mp}
We consider the Bethe free energy as the energy function of the interacting system of variables $\rho_{i\partial i}, \rho_{ij}$ and $\nu_{ij}\equiv \{ \nu_{i\to j}, \nu_{j\to i}\}$. Then an optimization algorithm can be obtained by studying the following statistical physics problem within a higher-level Bethe approximation,
\begin{equation}
\mathcal{Z}\equiv \sum_{\{\rho_{i\partial i}\}}\sum_{\{\rho_{ij}\}}\sum_{\{\nu_{ij}\}}e^{-\beta_{opt} F_{Bethe}}\prod_{i}\prod_{j \in \partial i} \delta(\rho_{ij}-\mathrm{Tr}_{\setminus i,j} \rho_{i\partial i})\delta(\nu_{i\to j}-\hat{\nu}_{i\to j}),
\end{equation}
where $\hat{\nu}_{i\to j}$ denotes the approximate BP equations \ref{app-BP-1} and \ref{app-BP-2}. To ensure that the reduced density matrices are
Hermitian and positive semidefinite, we take $\rho_{i\partial i}=e^{-\tilde{H}_{i\partial i}}/Z_{i\partial i}$ and $\rho_{ij}=e^{-\tilde{H}_{ij}}/Z_{ij}$ introducing the local effective Hamiltonians $\tilde{H}_{i\partial i}$ and $\tilde{H}_{ij}$ characterized by the set of couplings $g_{i\partial i}$ and $g_{ij}$, respectively. One can, in general, write   $\tilde{H}_{ij}=-\sum_{a,b=0,x,y,z}g_{ij}^{ab}\sigma_i^a\sigma_j^b$, and similarly for the $\tilde{H}_{i\partial i}$. 
 
Here we resort to the Bethe approximation to compute the local marginals of the effective Hamiltonians. To this end, we need the cavity marginals $M_{i\to j}(g_{ij},\nu_{ij})$ which are recursively determined by the set of neighboring cavity marginals $\{M_{k\to i}(g_{ik},\nu_{ik})|k \in \partial i \setminus j\}$ considering the local free energies and the local hard constraints \cite{MM-book-2009,R-prb-2012},
\begin{equation}\label{BP}
M_{i \to j}(g_{ij},\nu_{ij}) \propto \sum_{g_{i\partial i},\{g_{ik},\nu_{ik} | k \in \partial i \setminus j\}} \mathbb{I}_i e^{ -\beta_{opt}(\langle H_{i}\rangle-T \Delta s_i)}
\prod_{k \in \partial i \setminus j}\left(e^{-\beta_{opt}(\langle H_{ik}\rangle+T\Delta s_{ik})} M_{k \to i}(g_{ik},\nu_{ik}) \right),
\end{equation}
where for brevity we defined the indicator function $\mathbb{I}_i\equiv \prod_{k \in \partial i} \delta(\rho_{ik}-\mathrm{Tr}_{\setminus i,k} \rho_{i\partial i})\delta(\nu_{i\to k}-\hat{\nu}_{i\to k})$.
We are actually interested in the limit $\beta_{opt} \to \infty$, where the probability measure of the variables is concentrated on the optimal variable configuration(s). Taking the scaling $M_{i\to j}=e^{-\beta_{opt}\mathcal{M}_{i\to j}}$, we obtain the so called minsum equations \cite{KFL-inform-2001,BMZ-rsa-2005},
\begin{multline}\label{minsum}
\mathcal{M}_{i \to j}(g_{ij},\nu_{ij})=\min_{g_{i\partial i},\{g_{ik},\nu_{ik} | k \in \partial i \setminus j\}: \mathbb{I}_i}\left\{ \langle H_{i}\rangle-T \Delta s_i +
\sum_{k \in \partial i \setminus j}\left( \langle H_{ik}\rangle+T\Delta s_{ik}+ \mathcal{M}_{k \to i}(g_{ik},v_{ik})\right) \right\}.
\end{multline}
Note that the right hand side is computed conditioned on the constraints in $\mathbb{I}_i$. 
The equations can be solved by iteration starting from random initial messages $\mathcal{M}_{i \to j}(g_{ij},\nu_{ij})$ and updating them according to the above equations. After each update, we shift the minsum messages by a constant to keep $\min_{g_{ij},\nu_{ij}} \mathcal{M}_{i \to j}(g_{ij},\nu_{ij})=0$. Finally, one estimates the optimal couplings by minimizing the local minsum weights,
\begin{equation}
\mathcal{M}_i(g_{i\partial i},\{g_{ij},\nu_{ij}|j\in \partial i\})=\langle H_{i}\rangle-T \Delta s_i+
\sum_{j \in \partial i}\left( \langle H_{ij}\rangle+T\Delta s_{ij}+ \mathcal{M}_{j \to i}(g_{ij},\nu_{ij})\right).
\end{equation}

In practice, to implement the above algorithm, we have to work with discrete variables. The time complexity of the algorithm grows as $N_{b}^{ck_{max}^2}$ considering only the two-spin interactions in the $\tilde{H}_{i\partial i}$. Here, $N_{b}$ is the maximum number of bins in discrete representation of the variables, $k_{max}=\max_i |\partial i|$, and $c$ is a constant. Note that to update $\mathcal{M}_{i\to j}$, one only needs to sample over the $g_{i\partial i}$ and the incoming messages $\nu_{k \to i}$ as the $g_{ij}$ and the outgoing messages $\nu_{i\to k}$ are determined by the local hard constraints.


\begin{thebibliography}{prsty}


\bibitem{MP-epjb-2001} M. M\'ezard and G. Parisi,  Eur. Phys. J. B {\bf 20}, 217, 2001.

\bibitem{MM-book-2009} M. M\'ezard and A. Montanari, {\it Information, Physics, and Computation} (Oxford University Press, Oxford, 2009).

\bibitem{MZ-pre-2002}M. M\'ezard and R. Zecchina, Phys. Rev. E \textbf{66}, 056126 (2002).

\bibitem{MPZ-science-2002}M. M\'ezard, G. Parisi and R. Zecchina, Science \textbf{297}, 812 (2002).

\bibitem{KMRSZ-pnas-2007} F. Krzakala, A. Montanari, F. Ricci-Tersenghi, G. Semerjian and L. Zdeborov\'a,   Proc. Natl. Acad. Sci. {\bf 104}, 10318 (2007).


\bibitem{P-jphysa-2005} A. Pelizzola, J. Phys. A: Math. Gen. \textbf{38}:R309 (2005).

\bibitem{YFW-nips-2001}J. S. Yedidia, W. T. Freeman, and Y. Weiss, 
IEEE Trans. Infor. Theory \textbf{51}(7), 2282 (2005).


\bibitem{morita} T. Morita, J. Phys. Soc. Jpn. \textbf{12}(10), 1060 (1957).

\bibitem{CWW-prb-1992}L. De Cesare, K. Lukierska Walasek, and K. Walasek, Phys. Rev. B  \textbf{45}, 8127 (1992).


\bibitem{H-prb-2007}M. B. Hastings,  Phys. Rev. B \textbf{76}, 201102 (2007).

\bibitem{LP-aphys-2008}M. Lifer and D. Poulin,  Ann. Phys. (Leipzig) \textbf{323}, 1899 (2008).

\bibitem{LSS-prb-2008}C. Laumann, A. Scardicchio, and S. L. Sondhi,  Phys. Rev. B \textbf{78}, 134424 (2008).

\bibitem{KRSZ-prb-2008}F. Krzakala, A. Rosso, G. Semerjian, and F. Zamponi,  Phys. Rev. B \textbf{78}, 134428 (2008).

\bibitem{IM-prl-2010}L. B. Ioffe and M. M\'ezard, Phys. Rev. Lett.  \textbf{105}, 037001 (2010).


\bibitem{R-prb-2012} A. Ramezanpour, Phys. Rev. B \textbf{85}, 125131 (2012).


\bibitem{BFKSZ-pr-2013}V. Bapst, L. Foini,  F. Krzakala, G. Semerjian, and F. Zamponi,  Physics Reports \textbf{523}, 127 (2013).


\bibitem{DM-jstat-2010}O. Dimitrova and M. M\'ezard, 2011 J. Stat. Mech. P01020.


\bibitem{RZ-prb-2012}A. Ramezanpour and R. Zecchina, Phys. Rev. B \textbf{86}, 155147 (2012).

\bibitem{BR-jstat-2013} I. Biazzo and A. Ramezanpour, J. Stat. Mech. (2013) P04011.


\bibitem{PH-prl-2011}D. Poulin and M. B. Hastings,  Phys. Rev. Lett.  \textbf{106}, 080403 (2011).


\bibitem{ZV-prl-2004}M. Zwolak and G. Vidal, Phys. Rev. Lett. \textbf{93}, 207205 (2004).

\bibitem{VGC-prl-2004}F. Verstraete, J. J. Garcia-Ripoll, and J. I. Cirac, Phys. Rev. Lett. \textbf{93}, 207204 (2004).

\bibitem{FW-prb-2005}
A. E. Feiguin and S. R. White, Phys. Rev. B \textbf{72}, 220401 (2005).


\bibitem{KFL-inform-2001}F. R. Kschischang, B. J. Frey, and H. -A. Loeliger, IEEE Trans. Infor. Theory \textbf{47}, 498 (2001)


\bibitem{BMZ-rsa-2005}A. Braunstein, M. Mezard, and R. Zecchina, Random Structures and Algorithms  \textbf{27}, 201 (2005).


\end{thebibliography}
\end{document}